\documentclass[10pt]{article}
\usepackage{a4wide}                      % to get a wider page margen
\usepackage{epsf}                        % to make figures
\usepackage{latexsym}                    % to make fancy symbols
\usepackage{amsfonts}                    % to make fancy symbols
\usepackage{amssymb}                     % to make fancy symbols
\usepackage{cite}

\def\be{\begin{equation}}
\def\ee{\end{equation}}
\def\bea{\begin{eqnarray}}
\def\eea{\end{eqnarray}}

\def\nnw{\nonumber \\ [.2cm]}
\def\vsp#1{\vspace*{#1}}
\def\hsp#1{\hspace*{#1}}

\def\part{\partial}
\def\tfrac#1#2{{\textstyle{\frac{#1}{#2}}}}
\def\half{\tfrac{1}{2}}
\def\x{\times}

\def\cA{{\cal A}}
\def\cB{{\cal B}}
\def\cF{{\cal F}}
\def\cT{{\cal T}}

\def\cL{{\cal L}}  
\def\sqrtg{\sqrt{|g|}}

\def\R{\ensuremath{\mathbb{R}}}

\def\bGamma{\bar{\Gamma}}
\def\bR{\bar R} 
\def\bT{\bar T}
\def\bnabla{\bar \nabla}   

\def\mn{{\mu\nu}}
\def\mnr{{\mu\nu\rho}}

\def\makeatletter{\catcode`\@=11}% 11:letter
\makeatletter
\def\mathbox#1{\hbox{$\m@th#1$}}%
\def\math@ccstyles#1#2#3#4#5#6#7{{\leavevmode
      \setbox0\mathbox{#6#7}%
      \setbox2\mathbox{#4#5}%
      \dimen@ #3%
      \baselineskip\z@\lineskiplimit#1\lineskip\z@
      \vbox{\ialign{##\crcr
             \hfil \kern #2\box2 \hfil\crcr
             \noalign{\kern\dimen@}%
             \hfil\box0\hfil\crcr}}}}
\def\mathaccstyles{\math@ccstyles\maxdimen}
\def\maththroughstyles{\math@ccstyles{-\maxdimen}}
\def\unity%
 {\maththroughstyles{.45\ht0}\z@\displaystyle {\mathchar"006C}\displaystyle 1}
\begin{document}

\rightline{UG-FT-321/16}
\rightline{CAFPE-191/16}
%\rightline{XXYY/yy-nn}
\rightline{\today}
{~}
\vspace{1.4truecm}

%%%%%%%%%%%%%%%%%
\centerline{\LARGE \bf On the (non-)uniqueness of the Levi-Civita solution  }
\vspace{.3truecm}
\centerline{\LARGE \bf in the Einstein-Hilbert-Palatini formalism}
\vspace{1.3truecm}

\centerline{
    {\large \bf Antonio N. Bernal,${}^{a}$ }%\footnote{E-mail address: 
                                  %{\tt bjanssen@ugr.es} },
    {\large \bf Bert Janssen,${}^{b,c}$}
    {\large \bf Alejandro Jim\'enez-Cano,${}^{b,c}$}%\footnote{E-mail address: 
                                 % {\tt bjanssen@ugr.es} 
          } 
\centerline{
    {\large \bf Jos\'e Alberto Orejuela,${}^{b,c}$ }%\footnote{E-mail address: 
                                  %{\tt other.author@bla.zz}
    {\large \bf Miguel S\'anchez,${}^{a}$ }%\footnote{E-mail address: 
                                  %{\tt other.author@bla.zz}
   %{\bf and} 
    {\bf and} 
    {\large \bf Pablo S\'anchez-Moreno${}^{d,e}$ }\footnote{E-mail addresses: 
                                 anberna@hotmail.com, bjanssen@ugr.es, ajcfisica@correo.ugr.es, 
                                 josealberto@ugr.es, sanchezm@ugr.es, pablos@ugr.es
}                      
                                                            }
\vspace{.4cm}
\centerline{{\it ${}^a$Departamento de Geometr\'{\i}a y Topolog\'{\i}a}}
\centerline{{\it ${}^b$Departamento de F\'{\i}sica Te\'orica y del Cosmos}}
\centerline{{\it ${}^c$Centro Andaluz de F\'{\i}sica de Part\'{\i}culas Elementales}}
\centerline{{\it ${}^d$Departamento de Matem\'atica Aplicada}}
\centerline{{\it ${}^e$Instituto Carlos I de F\'{\i}sica Te\'orica y Computacional}}
\vsp{.2cm}

\centerline{{\it Facultad de Ciencias, Avda Fuentenueva s/n,}}
\centerline{{\it Universidad de Granada, 18071 Granada, Spain}}

\vspace{2truecm}

%%%%%%%%%%%%%%%%%
\centerline{\bf ABSTRACT}
\vspace{.5truecm}

\noindent
We study the most general solution for affine connections that are compatible with the 
variational principle in the Palatini formalism for the Einstein-Hilbert action (with
possible minimally coupled matter terms). We find that there is a family of solutions   
generalising the Levi-Civita connection, characterised by an arbitrary, non-dynamical 
vector field $\cA_\mu$. We discuss the mathematical properties and the physical implications
of this family and argue that, although there is a clear mathematical difference between
these new Palatini connections and the Levi-Civita one, both unparametrised geodesics and 
the Einstein equation are shared by all of them. Moreover, the Palatini connections are
characterised precisely by these two properties, as well as by other properties of its
parallel transport. Based on this, we conclude that physical
effects associated to the choice of one or the other will not be distinguishable, at least
not at the level of solutions or test particle dynamics. We propose a geometrical
interpretation for the existence and unobservability of the new solutions.

%%end of title page
\newpage
%%%%%%%%%%%%%%%%%%%%%%%%%%%%%%%%%%%%%%%%%%%%%%%%%%%%%%%%%
\noindent
\section{Introduction}
 
In the standard picture of General Relativity, gravitational physics is interpreted as 
physics occurring in a pseudo-Riemannian spacetime. From a mathematical point of view, spacetime 
is described as a 
$D$-dimensional,\footnote{Though for most physically relevant applications, $D$ is taken to be 4,
                we will work in arbitrary number of dimensions $D\geq 3$.} 
time-orientable Lorentzian manifold, equipped with a metric $g_\mn$ and its corresponding 
Levi-Civita connection, 
\be
\Gamma_\mn^\rho \ = \ \{_\mn^\rho\}
            \ \equiv \ \half \, g^{\rho\lambda} \Bigl(\partial_\mu g_{\lambda\nu} 
               \ + \ \partial_\nu g_{\mu\lambda} \ - \ \partial_\lambda g_{\mu\nu} \Bigr).
\label{Levi-Civita} 
\ee
This connection is defined as the unique connection that is both torsionless and metric 
compatible, %encoded by the conditions
\be
T_\mn^\rho \ \equiv \ \Gamma_\mn^\rho -   \Gamma_{\nu\mu}^\rho \ = \ 0, \hspace{2cm}
\nabla_\mu g_{\nu\rho} \ = \ 0,
\label{LCcond}
\ee
where $\nabla$ denotes the covariant derivative with respect to $\Gamma_\mn^\rho$.
The metric, in its turn, is a dynamical quantity, as it obeys the Einstein equations,
\be
R_\mn(g) \ - \ \half \, g_\mn R(g) \ = \ -\kappa \, \cT_\mn,  
\label{Einsteineqn0}
\ee
a set of second order differential equations for $g_\mn$, which can be derived through a variational
principle from the so-called Einstein-Hilbert action, minimally coupled to matter,
\be
S \ = \ \int d^Dx \, \sqrtg \, \Bigl[ \, \frac{1}{2\kappa} \, g^\mn R_\mn(g) 
                                  \ + \ {\cal L}_M (\phi, g) \Bigr]. 
\label{EHaction0}
\ee
In these equations, $R_\mn(g)$ is the Ricci tensor of the metric $g_\mn$, $R(g)$ the Ricci scalar, 
${\cal L}_M(\phi, g)$ the minimally coupled matter Lagrangian and $\cT_\mn$ its energy-momentum 
tensor. In a given spacetime, characterised by a metric $g_\mn$ which is a solution of 
(\ref{Einsteineqn0}) for a given $\cT_\mn$, free test particles will follow 
geodesic curves, described by the geodesic equation
\be
\ddot x^\mu \ + \ \{_{\nu\rho}^\mu\} \, \dot x^\nu \, \dot x^\rho \ = \ 0,
\label{LCgeodesic}
\ee
where $\dot x^\mu \equiv dx^\mu(\tau)/d\tau$ denotes derivation with respect to the proper time 
$\tau$ of the test particle. In this set-up, the metric components $g_\mn$ are the only 
gravitational degrees 
of freedom of the theory, as parallel transport, and hence also the curvature tensors, are 
completely determined by the metric through the Levi-Civita connection (\ref{Levi-Civita}). 
Traditionally, differential geometry in manifolds equipped with the Levi-Civita connection
is referred to as (pseudo-)Riemannian geometry.

However, in general in differential geometry, the metric and the affine connection are two 
independent quantities, that in principle play two different roles. The metric defines distances 
between points in the manifold and angles between vectors in the tangent space, while the affine 
connection provides a way of performing parallel transport of vectors and tensors along curves 
and hence defines the intrinsic curvature of the manifold. Only when the connection is chosen 
to be Levi-Civita (\ref{Levi-Civita}), both properties are fully determined by the metric, which
becomes the only dynamical quantity in the theory.

One could therefore ask whether there is a reason for the privileged status of the Levi-Civita 
connection in standard General Relativity and whether other choices for the connection are 
consistent and/or physically relevant. 

There are clear mathematical reasons to choose the Levi-Civita 
connection. Absence of torsion and metric compatibility  (\ref{LCcond}) are attractive 
mathematical features, which tend to simplify tensor identities considerably. Furthermore, 
the fact that the Levi-Civita connection is the only connection that combines these two 
properties yields it some kind of preferred status. 

At first sight, there are also physical reasons that seem to justify this choice of
connection. The Equivalence Principle, the cornerstone of General Relativity, which 
states that the gravitational force can be locally gauged away by a convenient choice of 
coordinates, is sometimes  summarised mathematically as the property that, at any point $p$ 
of the manifold, coordinates can be found such that the affine connection in that point 
vanishes,\footnote{See, for example, the discussion in \cite{We}, p. 74-75, which uses 
   formula (3.3.7) and, thus, (taking into account the version of the principle of equivalence 
   in p. 74), arrives at a symmetric connection;  however, other  approaches allow the
   existence of torsion explicitly, see for example  \cite[Ch. 4]{Hehl0}} 
$\Gamma_{\mn}^\rho(p) = 0$. However, it is clear that due to the tensorial character of the 
non-metric part of the connection $K_{\mn}^\rho = \Gamma_{\mn}^\rho - \{_{\mn}^\rho\}$, this 
property can only be accomplished if  $K_{\mn}^\rho$ vanishes identically.

Another feature of non-Levi-Civita connections is that affine geodesics and metric geodesics 
do not (necessarily) coincide and, since both types of geodesics have different mathematical 
meanings, general connections might give rise to potential difficulties as it comes to their
physical interpretation. Affine geodesics describe the straightest possible lines in a given 
geometry and represent the trajectory of unaccelerated particles (particles with covariantly 
constant four-velocities). On the other hand, metric geodesics describe the critical curve
between two points (in the timelike case, locally longest for the proper time)
and can easily be related to the trajectories of minimal action in absence 
of external forces. If both curves do not coincide, it is not clear which trajectory to 
adscribe to a free particle, but choosing the Levi-Civita connection the problem disappears 
naturally.

As convincing as some of these arguments might sound, the Levi-Civita connection   
(\ref{Levi-Civita}) still seems to appear as a convenient choice, not as a necessary tool. 
It would therefore be nice if there was a more rigorous, mathematical procedure that selects 
the Levi-Civita connection amongst other potential candidates. 

Such a procedure does in fact exist and is called the Palatini
formalism\footnote{As stressed in \cite[p. 23]{Hehl0}, such a name is unfortunate,  
       recall  \cite{FFR, Ei}.} 
\cite{Palatini}
(as opposed to the 
metric formalism, which simply assumes the Levi-Civita connection from the beginning). In the 
Palatini formalism, the connection is assumed to be a general affine connection 
$\Gamma_{\mn}^\rho$ and hence independent of the metric. The starting point of the 
Einstein-Hilbert-Palatini theory is then the Einstein-Hilbert action (\ref{EHaction0}), 
where now the Ricci tensor is written purely in terms of the general connection.
On the one hand, the Euler-Lagrange equation for the metric yields the Einstein equation, 
though in terms of a yet unknown connection, while on the other hand the Palatini equation,
the equation of motion
for the $\Gamma_{\mn}^\rho$, imposes conditions on the connection, which are clearly compatible 
with the Levi-Civita connection. The Levi-Civita connection arises thus in the Palatini 
formalism, not as a mere choice, but as a solution to the equations of motion, obtained from
a variational principle, much in the same way as the Einstein equation.

The Palatini formalism has been widely studied in different contexts, such as $f(R)$-gravity, 
Ricci-squared gravities and other extensions of standard General Relativity. For general 
Lagrangians, the Palatini formalism usually admits connections other than Levi-Civita, 
with different physics, which might yield alternatives to dark matter and/or dark 
energy or resolution of singularities \cite{CMQ, Querrella, ABFO, SL, TU, LBM, IKPP, BD, 
CDV, Bauer, CdL, Olmo}. On the other hand, it has also been proven 
\cite{ESJ, BJB, DP} that within the class of gravity theories with Lagrangians of the form 
$\cL (g_\mn, R_\mnr{}^\lambda)$ (i.e.~Lagrangians that are functionals of metric and the 
curvature tensors, but not of its derivatives), the Palatini formalism  yields the 
Levi-Civita connection as a solution only for those Lagrangians that are Lovelock gravities 
(and their equivalent Palatini counterparts). In other words, for Lovelock gravities, 
metric formalism is a consistent truncation of the Palatini formalism.  

It is sometimes claimed that the Levi-Civita connection is the only solution of the Palatini 
formalism, at least for the Einstein-Hilbert action. However, this assertion assumes implicitly 
either the symmetry or the metric compatibility of the connection. In fact, it has been known 
\cite{JS} (though it is often overlooked) that the Einstein-Hilbert action is invariant under 
the projective symmetry
\be
\Gamma_\mn^\rho \ \rightarrow \ \Gamma_\mn^{\prime\rho} 
                 \ = \ \Gamma_\mn^\rho \ + \ \cA_\mu\, \delta_\nu^\rho,
\ee
for an arbitrary vector field $\cA_\mu$, yielding the latter a gauge character. Yet it is 
\cite{Pons} that deals with the problem we are interested in: the 
``traditional'' Palatini problem of finding the most general connection allowed by the 
variational principle of the Einstein-Hilbert action and its physical and mathematical 
properties. 

As shown in \cite{Pons}, the most general solution for the Palatini equation of the 
Einstein-Hilbert action is given by a family of connections we will refer to as the 
``Palatini connections'',\footnote{For completeness, we point out that the Palatini 
             connection (\ref{Palatino0}) has also been studied in  \cite{TW, Ortin}, 
             though in different contexts and  with different conclusions to ours. Indeed, 
             in \cite{TW} a specific quadratic curvature term is added to the Einstein-Hilbert 
             term, which induces a particular, non-trivial dynamics for the connection, 
             while in \cite{Ortin} the Palatini connection is introduced as 
             an auxiliary field in the context of non-symmetric gravity theories. See also
             \cite{Deser} for relevant results in the context of Massive Topological Gravity. }
\be
\bGamma_\mn^\rho \ = \ \{_\mn^\rho\} \ + \ \cA_\mu\, \delta_\nu^\rho.
\label{Palatino0}
\ee
It is clear that the Palatini connections include the Levi-Civita
connection as a special case, but are generically non-metric compatible and non-symmetric. 
From the physical point of view, the connections contain a non-dynamical degree of freedom 
$\cA_\mu$, but it can easily be shown \cite{Pons} that the (symmetric part of) the Ricci 
tensor and hence the Einstein equation are not affected by the presence of this vector field.  
Based in this property, the authors of \cite{Pons} argue the vector field $\cA_\mu$ is 
undetectable and hence that the Palatini and the Levi-Civita connections are indistinguishable 
from a physical point of view. 

Even though we agree with most of interpretation of \cite{Pons}, we think that the invariance 
of the Einstein equation alone is insufficient to prove the undetectability of $\cA_\mu$. 
Indeed, each member of the Palatini family not only provides an Einstein equation, but also 
a specific parallel transport and its corresponding geodesics. As timelike and null geodesics
are the trajectories of free-falling test particles, any difference between Levi-Civita and
Palatini geodesics would be physically observable. Hence, the physical indistinguishability
of the Palatini connection from the Levi-Civita one can only be claimed if one succeeds in 
proving that the (timelike and null) geodesics of both connections coincide.
  
The aim of this paper is to show that the Palatini connections are indeed physically 
indistinguishable from Levi-Civita, not only in their Einstein equation, but also in their 
geodesics. As we will see, the parallel transport of the new Palatini solutions is different 
to the Levi-Civita one, but they differ only in a path-dependent homothety. Precisely
this homothety makes that the Palatini and the Levi-Civita connections share the same 
pre-geodesics, i.e.~they have the same geodesics, up to reparametrisations. As we will show, 
this property is unique for the Palatini connections and provides a strong support to the 
claimed undetectability. At the same time, the homothetic difference between the geodesics 
of both connections suggests an interpretation for the projective symmetry of the 
Einstein-Hilbert action and the Einstein equations observed on \cite{JS, Pons} as an
unphysical degree of freedom, related to the freedom of reparametrisation of the geodesics.
  
The organisation of this paper is as follows: in Section \ref{solution} we derive the
equations of motion of the Einstein-Hilbert-Palatini theory and,  for the sake of completeness, 
we deduce the most general solution for the Palatini equation, generically non-metric 
compatible and non-symmetric. In Section \ref{properties} we discuss the geometrical
properties of the solutions, pointing out the mathematical differences between the Palatini
and the Levi-Civita connection. In Section \ref{interpretation}, we argue that the Palatini
connections have no observable effects on the physics of solutions and test particle dynamics
and hence turn out to be indistinguishable from the Levi-Civita connection. Finally in Section
\ref{conclusion} we propose a physical interpretation of the Palatini connections and
elaborate on what remains of the preferred status of the Levi-Civita connection in General
Relativity.

%%%%%%%%%%%%%%%%%%%%%%%%%%%%%%%%%%%%%%%%%%%%%%%%%%%%%%%%
\section{The solution}
\label{solution}

Consider the $D$-dimensional Einstein-Hilbert action in the Palatini formalism, minimally 
coupled to a generic matter field $\phi$, 
\be
S (g, \Gamma) \ = \ \int d^Dx \sqrtg \, 
         \Bigl[\, \tfrac{1}{2\kappa}\,g^\mn R_\mn (\Gamma) \ + \ \cL_M(\phi, g)\Bigr]
\label{EHaction}
\ee
where the metric $g_\mn$ and the connection $\Gamma_\mn^\rho$ are treated as independent 
variables. We assume the connection to be completely general, without imposing neither symmetry, 
nor metric compatibility, such that the Ricci tensor, in our conventions given by
\be   
R_\mn (\Gamma) \ \equiv \ R_{\mu\lambda\nu}{}^\lambda   (\Gamma)
                 \ = \ \partial_\mu \Gamma_{\lambda\nu}^\lambda 
                 \ - \ \partial_\lambda \Gamma_{\mn}^\lambda
                 \ + \ \Gamma_{\mu\rho}^\lambda \, \Gamma_{\lambda\nu}^\rho 
                 \ - \ \Gamma_{\lambda\rho}^\lambda \, \Gamma_{\mu\nu}^\rho,
\ee
is completely independent of the metric. Note that the action (\ref{EHaction}) is first order 
in the connection and zeroth order in the metric  (in contrast to the metric formalism, 
where (\ref{EHaction0}) is second order in $g_\mn$). In the Palatini formalism it is therefore 
not necessary to include a Gibbons-Hawking-York term \cite{York, GH}, as there are no boundary 
terms coming from the variation of second order terms.

The Palatini formalism prescribes that the physics of the above action is given by the 
Euler-Lagrange equations of the metric, the connection and the matter fields. However, as 
we assume the matter Lagrangian to be minimally coupled, the matter equations of motion 
do not couple to the connection  and hence, for the purposes we are interested in, in this 
letter, the matter sector will not play any relevant role. Except for its energy-momentum 
tensor in the Einstein equation, we will omit all references to the matter fields from now on.  

The Einstein equation, the 
variation of the action with respect to the metric, is given by
\be
0 \ = \ \frac{2\kappa}{\sqrtg} \, \frac{\delta S}{\delta g^\mn} \ = \ 
R_{(\mn)} (\Gamma) \ - \ \half g_\mn\, R(\Gamma) \ + \ \kappa \, {\cal T}_\mn,
\label{Einsteineqn}
\ee 
where $R_{(\mn)}$ indicates the symmetric part of the Ricci tensor. On the other hand, the 
variation of the action (\ref{EHaction}) with respect to the connection can be easily done 
by first computing the Palatini Identity, the variation of the Ricci tensor with respect to 
the connection,
\be
\delta R_\mn (\Gamma) \ = \ \nabla_\mu(\delta \Gamma_{\lambda\nu}^\lambda) 
                          \ - \ \nabla_\lambda(\delta \Gamma_{\mu\nu}^\lambda) 
                          \ + \ T_{\mu\lambda}^\rho (\delta \Gamma_{\rho\nu}^\lambda),
\ee
where we use $\nabla$ and $T_\mn^\rho$ to denote the covariant derivative and the torsion 
associated to the connection $\Gamma_\mn^\rho$ respectively. The variation of (\ref{EHaction}) 
is obtained by substituting the Palatini Identity and integrating by parts, yielding
the Palatini equation (compare with \cite{Hehl1})
\bea
&& \nabla_\lambda g^\mn \ - \ \nabla_\sigma g^{\sigma\nu} \, \delta^\mu_\lambda
   \ + \ \half g^{\rho\tau} \nabla_\lambda g_{\rho\tau}\,  g^\mn
        \ - \ \half g^{\rho\tau} \nabla_\sigma g_{\rho\tau}  \, g^{\sigma\nu}\, \delta^\mu_\lambda \nnw
&& \hsp{1cm} - \ T_{\rho\lambda}^\rho \, g^\mn  
     \ + \ T_{\rho\sigma}^\rho \, g^{\sigma\nu}  \, \delta^\mu_\lambda 
      \ + \ T_{\sigma\lambda}^\mu \, g^{\sigma\nu} \ = \ 0.
\label{Palatinieqn}
\eea

Both the Einstein equation (\ref{Einsteineqn}) and the Palatini equation (\ref{Palatinieqn}) can 
be simplified: substracting the trace of (\ref{Einsteineqn}) and the $\delta^\lambda_\mu$ and 
the $g_\mn$ traces of (\ref{Palatinieqn}), these equations reduce respectively to 
\bea
&& R_{(\mn)} (\Gamma) \ = \ -\kappa \Bigl[ \cT_\mn \ - \ \tfrac{1}{D-2}\, g_\mn \cT\Bigr], 
\label{tracelessEinstein}\\ [.3cm]
&& \nabla_\lambda g_\mn \ - \ T_{\nu\lambda}^\sigma \, g_{\sigma\mu}
                      \ - \ \frac{1}{D-1} \ T_{\sigma\lambda}^\sigma \, g_{\mn}
                      \ - \ \frac{1}{D-1} \ T_{\sigma\nu}^\sigma \, g_{\mu\lambda} \ = \ 0.
\label{tracelessPalatini}
\eea
The idea is now to solve the Palatini equation for $\Gamma_\mn^\rho$ and substitute this solution 
in the Einstein equation to determine the geometry of the spacetime. Note that the Palatini 
equation is not a dynamical equation for $\Gamma_\mn^\rho$, but just an algebraic constraint. 
This is due to the fact that there are no kinetic terms for the connection in the  
Einstein-Hilbert action, which in turn is intimately related to the fact that the (metric) 
Einstein-Hilbert action is the first order Lovelock 
Lagrangian \cite{BJB}.

It is trivial to see that the Levi-Civita connection (\ref{Levi-Civita}) is a solution of the 
Palatini equation, as each term in (\ref{tracelessPalatini}) is identically zero, due to the 
necessary conditions (\ref{LCcond}). It is also straightforward to see that 
(\ref{tracelessPalatini}) forces any symmetric 
connection to be metric-compatible and vice versa. Hence assuming any of the two conditions 
(\ref{LCcond}) is sufficient in the Einstein-Hilbert-Palatini formalism for the connection to be 
Levi-Civita, as the other one will be automatically imposed by the Palatini equation.
However, the question remains whether there exist non-symmetric and non-metric compatible 
connections that are solutions of (\ref{tracelessPalatini}).

The Palatini equation (\ref{tracelessPalatini}), being an algebraic equation, is easy to solve. 
In fact,
the general solution can be found in the same way as the expression  (\ref{Levi-Civita})
for the Levi-Civita connection is deduced from the conditions (\ref{LCcond}). Writing  
(\ref{tracelessPalatini}) explicitly in terms
of the connections and cyclically permuting the free indices, we find
\bea
&& \partial_\lambda g_{\mu\nu} \ - \ \Gamma_{\lambda\mu}^\sigma g_{\sigma\nu} 
                         \ - \ \Gamma_{\nu\lambda}^\sigma g_{\mu\sigma} 
                     \ - \ \tfrac{1}{D-1} \ T_{\sigma\lambda}^\sigma \, g_{\mn}
                      \ - \ \tfrac{1}{D-1} \ T_{\sigma\nu}^\sigma \, g_{\mu\lambda} \ = \ 0, \nnw
&& \partial_\mu g_{\nu\lambda} \ - \ \Gamma_{\mu\nu}^\sigma g_{\sigma\lambda} 
                         \ - \ \Gamma_{\lambda\mu}^\sigma g_{\nu\sigma} 
                     \ - \ \tfrac{1}{D-1} \ T_{\sigma\mu}^\sigma \, g_{\nu\lambda}
                      \ - \ \tfrac{1}{D-1} \ T_{\sigma\lambda}^\sigma \, g_{\nu\mu} \ = \ 0, \nnw
&& \partial_\nu g_{\lambda\mu} \ - \ \Gamma_{\nu\lambda}^\sigma g_{\sigma\mu} 
                         \ - \ \Gamma_{\mu\nu}^\sigma g_{\lambda\sigma} 
                     \ - \ \tfrac{1}{D-1} \ T_{\sigma\nu}^\sigma \, g_{\lambda\mu}
                      \ - \ \tfrac{1}{D-1} \ T_{\sigma\mu}^\sigma \, g_{\lambda\nu} \ = \ 0.
\eea
Adding up the last two equations and subtracting the first one, we find that the connection 
$\Gamma_\mn^\rho$ can be expressed in terms of the trace of its torsion and the Levi-Civita 
connection:
\be
\Gamma_\mn^\rho \ = \ \{_\mn^\rho\} \ - \ \tfrac{1}{D-1} \, T_{\sigma\mu}^\sigma \, \delta_\nu^\rho.
\ee
Using group-theoretical arguments, it is easy to see that the trace of the torsion can be fully 
represented by a $D$-dimensional vector, $T_{\sigma\mu}^\sigma =  -(D-1) \, \cA_\mu$. 
We conclude therefore that the most general solution of the Palatini equation 
(\ref{tracelessPalatini}) can be written in the form (see also \cite{JS, Pons, TW, Ortin}) 
\be
\Gamma_\mn^\rho \ = \  \bGamma_\mn^\rho \ \equiv \ \{_\mn^\rho\} \ + \ \cA_\mu \, \delta_\nu^\rho,
\label{Pconnection}
\ee
with $\cA_\mu$ an arbitrary, non-dynamical vector field. Note that the Levi-Civita connection is
trivially recovered, choosing $\cA_\mu=0$. From the construction it is clear 
that (\ref{Pconnection}) is indeed the most general solution to the Palatini equation 
(\ref{tracelessPalatini}).

%%%%%%%%%%%%%%%%%%%%%%%%%%%%%%%%%%%%%%%%%%%%%%%%%%%%%%%%
\section{Geometrical properties}
\label{properties}
 
Now that we have found the most general solution (\ref{Pconnection}) to the Palatini equation, 
we will study in this section its geometrical properties and give a physical interpretation in 
the next one. 

As we mentioned in the construction, the (non-trivial, {\it i.e.} non-Levi-Civita) Palatini 
connections (\ref{Pconnection}) are \textit{neither 
symmetric, nor metric compatible}, the generalisation of (\ref{LCcond}) being
\be
\bT_\mn^\rho \ = \  \cA_\mu \, \delta_\nu^\rho \ - \  \cA_\nu \, \delta_\mu^\rho,
\hsp{2cm}
\bnabla_\mu g_{\nu\rho} \ = \ - 2 \, \cA_\mu  \, g_{\nu\rho}.
\label{bGammacond}
\ee
The corresponding curvature tensors are given by 
\be
\bR_\mnr{}^\lambda \, = \,  R_\mnr{}^\lambda(g) \, + \, \cF_\mn (\cA) \, \delta_\rho^\lambda, \hsp{1cm} 
\bR_{\mu\nu} \, = \,  R_{\mu\nu}(g) \, + \, \cF_{\mu\nu}  (\cA),  \hsp{1cm}
\bR \ = \  R(g),
\label{barR}
\ee
where $R_\mnr{}^\lambda(g)$, $ R_{\mu\nu}(g)$ and $R(g)$ are respectively the Riemann tensor, the 
Ricci tensor and the Ricci scalar with respect to the Levi-Civita connection, 
$\bar R=g^{\mu\nu}\bR_{\mu\nu}$ the Ricci scalar associated to $\bar R_{\mu\nu}$ and 
$\cF_{\mu\nu}(\cA) = \partial_\mu \cA_\nu - \partial_\nu \cA_\mu$.\footnote{The way the vector
      field $\cA_\mu$ appears in the curvature tensors reminds strongly of the electromagnetic 
      field strength tensor $F_{\mu\nu} = \partial_\mu A_\nu - \partial_\nu A_\mu$ in Maxwell theory, 
      as the curvature tensors are 
      invariant under the transformation $\cA_\mu \rightarrow \cA_\mu + \partial_\mu \Lambda$. 
      This is the reason why $\cA_\mu$ is interpreted as a $U(1)$ gauge field in  \cite{TW},
      even though the torsion and the parallel transport are not. We will show that $\cA_\mu$
      is not a Maxwell-like gauge field, but that the whole of $\cF_{\mu\nu}$ is undetectable.} 
Note that the Riemann and Ricci 
tensors (\ref{barR}) do not satisfy the symmetry properties of their Levi-Civita counterparts, 
due to (\ref{bGammacond}). Yet it is interesting to notice that the symmetric part of the 
Ricci tensor, the one determined by the Einstein equations, coincides precisely with the Ricci 
tensor of the Levi-Civita connection: $\bR_{(\mu\nu)} \ = \  R_{\mu\nu}(g)$.
\vsp{.2cm}

A remarkable property of the Palatini connections (\ref{Pconnection}) is that \textit{affine 
geodesics turn out to be pregeodesics of the Levi-Civita connection} 
 ({\it i.e.}~they describe the same trajectories in the manifolds, though with a different 
parametrisation). Indeed, the affine geodesic equation for the Palatini connections, 
$\dot x^\rho \bnabla_\rho \dot x^\mu = 0$, written in terms of $\{_\mn^\rho\}$ and $\cA_\mu$, 
take the form 
\be
\dot x^\rho \nabla^{(g)}_\rho \dot x^\mu \ = \ -  \cA_\rho \, \dot x^\rho \, \dot x^\mu, 
\label{bGammageodesic}
\ee  
where $\nabla^{(g)}$ denotes the covariant derivative with respect to the Levi-Civita connection. 
The equation of all (non-lightlike) pregeodesics can be derived as an extremum of the arc 
length functional
\be
s(\lambda) \ = \ \int_0^\lambda \sqrt{|g_\mn\, \dot x^\mu \, \dot x^\nu\, |} \,  d\lambda',
\label{funcional}
\ee
where $\dot x^\mu \equiv dx^\mu(\lambda')/d\lambda'$ denotes derivation with respect to an 
arbitrary parameter $\lambda'$. Extrema of this functional in general take the form
\be
\dot x^\rho \nabla^{(g)}_\rho \dot x^\mu 
%\ddot x^\mu \ + \ \{_{\nu\rho}^\mu\} \, \dot x^\nu \, \dot x^\rho 
               \ = \ \Bigl(\frac{\ddot s}{\dot s}\Bigr) \, \dot x^\mu,
\label{LCgeodesic2}
\ee
but the equation (\ref{bGammageodesic}) can be recovered with the specific parameter choice
\be
s(\lambda) \ = \ \int_0^\lambda e^{-G(\lambda')}\,  d\lambda' 
       \hsp{1cm} \mbox{with}  \hsp{.5cm} 
     G(\lambda) \ = \ \int_0^\lambda \dot x^\rho \cA_\rho \, d\lambda'. 
\label{G(lambda)}
\ee
This observation proves the two points mentioned above: first of all that (\ref{bGammageodesic})
can be interpreted as both the equation of the geodesics of $\bar \nabla$ and the equation of 
a particular type of reparametrisations of the Levi-Civita geodesics.
Secondly, that the right-hand side of (\ref{bGammageodesic}) can be absorbed 
in a conveniently chosen (though geodesic-dependent) reparametrisation of the geodesics. In 
particular, (\ref{bGammageodesic}) can be transformed into the geodesic equation of the 
Levi-Civita connection (\ref{LCgeodesic}) through the change of parameter
\be
\frac{dx^\mu(\lambda)}{d\lambda} \ = \ \frac{dx^\mu(\tau)}{d\tau} \, \frac{d\tau}{d\lambda}
\hsp{1cm} \mbox{with}  \hsp{.5cm}  \frac{d\tau}{d\lambda} \ = \ e^{-G(\lambda)}.
\label{reparametrisation}
\ee
Summing up, the curves described by (\ref{LCgeodesic}) and by  (\ref{bGammageodesic}) 
yield the trajectories of the geodesics, with different parametrisations controlled by 
(\ref{reparametrisation}).

The Palatini connections (\ref{Pconnection}) are not the only connections that have the same 
pregeodesics as $\{_\mn^\rho\}$.  Indeed, affine connections with the same pregeodesics are
called \textit{projectively related}
and any affine connection projectively related to Levi-Civita's has the form
$\tilde \Gamma_\mn^\rho =  \{_\mn^\rho\}  + \cA_\mu \delta^\rho_\nu 
            +  \cB_\nu \delta^\rho_\mu$ \cite{Spivak}. 
However it is interesting to notice that the curvature tensors coming from this connection 
have more complicated expressions than the ones given in 
(\ref{barR}). For example, the Riemann tensor associated with  
$\tilde \Gamma_\mn^\rho$ is given by 
\be
\tilde R_\mnr{}^\lambda \, = \, R_\mnr{}^\lambda(g) \, + \, \cF_\mn(\cA) \, \delta_\rho^\lambda
\, + \, (\nabla_\mu \cB_\rho \, - \, \cB_\mu \cB_\rho)\, \delta_\nu^\lambda
\, + \, (\nabla_\nu \cB_\rho \, - \, \cB_\nu \cB_\rho)\, \delta_\mu^\lambda.
\ee
Notice in particular, that the symmetric part is of the Ricci tensor in general does not coincide
with the Levi-Civita Ricci 
tensor.\footnote{\label{fn}It does if and only if $\cB_\mu$ satisfies the condition
       $\nabla_\mu \cB_\nu \, = \, \cB_\mu \cB_\nu$. This a the well-known condition of
       recurrence, see \cite{Stephani} or \cite{Hall}. However this particular condition 
       breaks the generic character of the projectively related connections. Indeed, it is
       not difficult to show that such a $\cB_\mu$ implies the existence of a parallel
       (covariantly constant) 1-form ${\cal P}_\mu$, pointwise proportional to  $\cB_\mu$.
       We will 
       briefly comment on this case later in this paper.}

\vsp{.2cm}

Since the Palatini geodesics are pregeodesics of the Levi-Civita ones, it should be clear 
that they have the same geodesic deviation (modulo the pointwise direction of the velocity
of the geodesic), as their trajectories in the spacetime manifold 
coincide. However this can also  be made explicit, starting from the geodesic deviation equation 
for the arbitrary connections (see for example \cite{SS}), applied to the Palatini connections 
(\ref{Pconnection}), 
\be
\tfrac{\partial x^\mu}{\partial\lambda} 
      \bnabla_\mu \Bigl[\tfrac{\partial x^\nu}{\partial\lambda} 
               \bnabla_\nu \tfrac{\partial x^\alpha}{\partial\eta}\Bigr]
\ + \ \bR_\mnr{}^\alpha \, \tfrac{\partial x^\mu}{\partial\eta} \, 
                         \tfrac{\partial x^\nu}{\partial\lambda} \, 
                         \tfrac{\partial x^\rho}{\partial\lambda}
\ - \ \tfrac{\partial x^\nu}{\partial\lambda}  
         \bnabla_\mu \Bigl[ \bT_{\nu\rho}^\alpha \, \tfrac{\partial x^\nu}{\partial\lambda}\, 
                              \tfrac{\partial x^\rho}{\partial\eta}  \Bigr] \ = \ 0, 
\label{Pgeoddev}
\ee
and seeing that its maps to the geodesic deviation equation for the Levi-Civita connection,
\be
\tfrac{\partial x^\mu}{\partial\tau} 
      \nabla_\mu \Bigl[ \tfrac{\partial x^\nu}{\partial\tau} 
              \nabla_\nu  \tfrac{\partial x^\alpha}{\partial\sigma}\Bigr]
\ + \ R_\mnr{}^\alpha \, \tfrac{\partial x^\mu}{\partial\sigma} \, 
                         \tfrac{\partial x^\nu}{\partial\tau} \, 
                         \tfrac{\partial x^\rho}{\partial\tau} \ = \ 0,
\label{LCgeoddev}
\ee
under the reparametrisation
\be
\frac{\partial x^\mu}{\partial\lambda} \ = \
           \frac{\partial \tau}{\partial\lambda}\, \frac{\partial x^\mu}{\partial\tau}, 
\hsp{2cm}
\frac{\partial x^\mu}{\partial\eta} \ = \  
     \frac{\partial \tau}{\partial\eta}\frac{\partial x^\mu}{\partial\tau} \ + \
     \frac{\partial x^\mu}{\partial\sigma}, 
\ee
where $\tau = \tau(\lambda, \eta)$ is defined as
\be
\tau(\lambda, \eta) \ = \ \int_0^\lambda  e^{-G(\lambda', \eta)} d\lambda' 
   \hsp{1cm} \mbox{with}  \hsp{.5cm} 
G(\lambda,\eta) \ = \  \int_0^\lambda \dot x^\rho \cA_\rho \, d\lambda'.
\ee
Hence, the Palatini and the Levi-Civita connections have the same geodesic deviation, as
solutions of (\ref{Pgeoddev}) are also solutions of (\ref{LCgeoddev}) and vice versa.

\vsp{.2cm}
Another remarkable property of the Palatini connections is that \textit{its parallel transport
  becomes 
homothetic with respect to the Levi-Civita connection.} From the very definition of the Palatini 
connections (\ref{Pconnection}), it is clear that the difference between parallel transport
of a vector $V^\mu$ along a curve $x^\mu = x^\mu(\lambda)$ according to $\bGamma_\mn^\rho$ and 
according to  $\{_\mn^\rho\}$ is proportional to the vector itself:
\be
\dot x^\rho \bnabla_\rho V^\mu \ - \ \dot x^\rho \nabla^{(g)}_\rho V^\mu 
     \ = \ \dot x^\rho \cA_\rho \, V^\mu.
\ee 
Concretely this means that the result of parallelly transporting vectors with the Levi-Civita
or with the Palatini connections leads to different results, but the resulting vectors only 
differ in their norm (or, more properly  for the lightlike case, in a proportionality coefficient).
Indeed, if $V_{g}^\mu (\lambda)$ is the result of parallel transport 
along a curve  $x^\mu = x^\mu(\lambda)$ according to $\{_\mn^\rho\}$, then the result of parallel 
transport along the same curve according to  $\bGamma_\mn^\rho$ is given by
\be
V_{\bGamma}^\mu (\lambda) \ = \  e^{-G(\lambda)} \, V_{g}^\mu(\lambda), 
\label{homotheticfactor}
\ee
with $G(\lambda)$ given by (\ref{G(lambda)}). Note that the proportionality coefficient depends 
on the curve  $x^\mu = x^\mu(\lambda)$, but not on $V^\mu$. We therefore have that Palatini 
transport is equal to Levi-Civita transport, composed by a homothety of ratio $ e^{-G(\lambda)}$.

The property of non-constant norm under parallel transport is a consequence of the fact that 
the Palatini connections are not metric compatible. A general feature of non-metric compatible 
connections is that parallel transport does not conserve the scalar product of vectors. 
For the Palatini transport of two vectors $V^\mu$ and $W^\mu$, we have that
\be
\dot x^\rho \bnabla_\rho (g_\mn V^\mu W^\nu) 
            \ = \  \dot x^\rho \bnabla_\rho g_\mn \, V^\mu W^\nu
           \ = \ 2 \, \dot x^\rho \cA_\rho \, V_\mu W^\mu 
           \ = \  2\, G'(\lambda) \, V_\mu W^\mu.
\ee  

The Palatini connections (\ref{Pconnection}) are the \textit{only connections yielding this
  homothety property under parallel transport for all vectors along any 
curve.}\footnote{It is worth pointing out that  Palatini connections present some formal 
       analogies with the {\em standard volume-preserving connections}, exhibited in  
       \cite[sections 3.10, 3.12]{Hehl2} as a natural example of connections arising from 
       a decomposition in ``Riemannian and post-Riemannian pieces''. However, ours are 
       allowed to have a non-volume preserving transport and a non-symmetric Ricci tensor.} 
Indeed, an arbitrary 
connection $\Gamma_\mn^\rho = \{_\mn^\rho\} + K_\mn^\rho$ with an arbitrary tensor $K_\mn^\rho$ 
yields homothetic parallel transport with respect to Levi-Civita if and only if
\be
\dot x^\nu K_{\nu\rho}^\mu V^\rho \ = \ f(\lambda) \, V^\mu,
\label{homothetic1}
\ee
for some function $f(\lambda)$, which may depend on the curve followed. The above expression 
is true for all vectors $V^\mu$, if and only if 
$\dot x^\nu K_{\nu\rho}^\mu = f(\lambda)\delta^\mu_\rho$.
The homothety condition (\ref{homothetic1}) can therefore be written 
as
\be
\dot x^\nu K_{\nu\rho}^\mu \ = \ \dot x^\nu \cA_{\nu} \, \delta^\mu_\rho.
\label{homothetic2}
\ee
Then it is easy to see that, in order for this identity to be true for all curves 
$x^\mu = x^\mu(\lambda)$, we must have that $K_{\nu\rho}^\mu = \cA_{\nu} \, \delta^\mu_\rho$ and 
hence that $\Gamma_\mn^\rho = \bGamma_\mn^\rho$. In particular, the Levi-Civita connection can be 
characterised as the unique symmetric connection such that its parallel transport becomes a 
metric homothety (which turns out trivial, {\it i.e.}, a isometry). 
As a last remark, notice also that to take only curves $x^\mu$ that are timelike is  
enough to prove all these characterisations regarding homotheties. 

%%%%%%%%%%%%%%%%%%%%%%%%%%%%%%%%%%%%%%%%%%%%%%%%%%%%%%%%
\section{Physical observability}
\label{interpretation}

Let us briefly summarise the main results of the previous sections. We found that the most 
general solution to the Palatini equation (\ref{tracelessPalatini}) is given by the family of
Palatini connections
 (\ref{Pconnection}),
\be
\bGamma_\mn^\rho \ = \ \{_\mn^\rho\} \ + \ \cA_\mu \, \delta_\nu^\rho,
\label{Pconnection2}
\ee
yielding curvature tensors (\ref{barR}) that only differ from the Levi-Civita tensor by terms
involving $\cF_\mn(\cA)$. In particular, the symmetric part of the Ricci tensor is identical to 
the Levi-Civita Ricci tensor,
\be
\bR_{(\mn)} \ = \ R_\mn(g).
\label{bR=R}
\ee

Furthermore, we have found that the Palatini connections (\ref{Pconnection2}) are unique in two 
ways:

\begin{itemize}
\item  $\bGamma_\mn^\rho$ are the only connections that, for a given metric $g_\mn$ have the 
same pregeodesics as $\{_\mn^\rho\}$ and at the same time satisfy the relation 
(\ref{bR=R}) between its Ricci tensor and 
$R_\mn(g)$.\footnote{With the exception of the connection $\tilde \Gamma_\mn^\rho = \{_\mn^\rho\}
  + \cB_\nu \delta_\mu^\lambda$ with $\nabla_\mu \cB_\nu = \cB_\mu \cB_\nu$, as explained in footnote
  \ref{fn}. These connections  appear only when there exists a parallel ${\cal P}_\mu$ and, then,
  one can write locally ${\cal P}_\mu = dx$ and $\cB_\mu = dx/(1-x)$. When  ${\cal P}_\mu$ is not
  lightlike then the spacetime decomposes as a product $N\x \R$ and $x$ can be regarded as a
  natural coordinate on $\R$. When it is lightlike, then the spacetime becomes a Brinkmann space
  (see \cite{BSS}) and $x$ can be regarded as a natural lightcone coordinate $u$. This shows the
  exceptionality of these connections ``physically indistinguishable to Levi-Civita, but not
  Palatini.'' \label{fn2}}

\item  $\bGamma_\mn^\rho$ are the only connections whose parallel transport of any vector
along timelike curves (and, then, along any curve) is homothetic to the Levi-Civita transport.   
\end{itemize}

Given that the Palatini connections are mathematically clearly different from the Levi-Civita 
connection, one wonders whether it would also lead to different physics and, in case it does,
whether (any of) these connections describe correctly our universe. The question is especially 
important in the light of the issue about the preferred status of the Levi-Civita 
connection in General Relativity. If the Palatini connections have physically observable 
effects, then the question remains why Levi-Civita is singled out, as there seems to be no 
experimental or observational evidence that supports the existence of a non-trivial vector 
field $\cA_\mu$. On the other hand, if the Palatini connections turn out to be physically 
indistinguishable from the Levi-Civita connection, then there seems to be a surprising 
``duality symmetry'' (rather than the commented $U(1)$ gauge invariance pointed out 
in \cite{TW}), that relates mathematically different spaces as physically equivalent. 

In our opinion, the first uniqueness property of the Palatini connections stated above, 
suggests the latter possibility, namely, the physical indistinguishability of all 
the Palatini connections at least in ``rough'' physics, i.e.~at the level of
the solutions of the Einstein equations and the dynamics of test particles.

First of all, the fact that the symmetric part of the Ricci tensor of the Palatini connection 
coincides exactly with the Ricci tensor of the Levi-Civita connection implies that the 
explicit form of the Einstein equation is independent of the choice of $\cA_\mu$: any 
metric $g_\mn$ that, for a given minimally coupled $\cT_\mn$, is a solution of the 
Einstein equations (\ref{tracelessEinstein}) with $\Gamma_\mn^\rho = \bGamma_\mn^\rho$, is also 
a solution of the same Einstein equations with $\Gamma_\mn^\rho = \{_\mn^\rho\}$, and 
vice versa. Furthermore, these solutions coincide with the solutions of the Einstein equation 
(\ref{Einsteineqn0}) in the metric formalism, which proves the complete equivalence of both 
formalisms at the level of the solutions.  

The invariance of the Einstein equation was already pointed out in \cite{Pons}. However, 
under our viewpoint, this property alone is not enough to ensure the undetectability of 
Palatini connections, as it does not take into account an important issue in  gravitational
physics: the dynamics of test particles. Notice that if the argument of \cite{Pons} were 
sufficient, one should consider as physically indistinguishable from Levi-Civita    
all the affine connections that leave the symmetric part of the Ricci tensor invariant, 
whether they are solutions to the Palatini equation (\ref{tracelessPalatini}), or not. 
However, such connections would have their own geodesics, which may be very different to 
Levi-Civita ones. A trivial counterexample in Minkowski spacetime would be the affine 
connection whose components $\hat\Gamma_\mn^\rho$ in natural coordinates all vanish except 
$\hat\Gamma_{00}^1=1$. Obviously, this connection is flat (and hence has the same Ricci 
tensor as the Levi-Civita connection). Yet, the curve $x^\mu (\lambda )=\lambda\, \delta^{\mu}_0$, 
being a geodesic for the Levi-Civita connection but not for $\hat\Gamma_\mn^\rho$,  
would represent two physically clearly distinguishable situations (free fall and accelerated
motion) depending of the choice of the connection.

It is therefore necessary to consider not only the invariance of the Einstein equations, but 
also the equivalence of the geodesic motion of both connections. As we have shown, this is 
indeed the case, thanks to the fact that the (timelike) geodesics of the Palatini connection 
are pregeodesics of the Levi-Civita connection. In other words, the spacetime trajectories of 
free-falling test particles for the Palatini connection are the same as the ones described 
for Levi-Civita. As the latter respect the Equivalence Principle, so does the former: for 
any of the two connections considered, the outcome of any local experiment in a free 
falling system will be independent of the velocity and the location of the system in 
spacetime. Furthermore, also all non-local effects, which show up in tidal forces, will 
be the same for both connections, as the pregeodesic deviation equations for the two cases 
are equivalent.  Hence we find that also at the level of the motion of test 
particles, the physics of the Palatini connections is indistinguishable from standard 
physics.\footnote{In this light, it can be argued that the mathematical formulation
of the Equivalence Principle should not be, as stated in the introduction, that the 
connection should be Levi-Civita, but rather that the connection should have the same 
pregeodesics as Levi-Civita.}

On the other hand, the second characterisation of the Palatini connections needs a  subtler 
analysis. One might argue that there must be physical effects that become visible in the 
parallel transport of vectors: as in general the results of parallel transport according 
to the Palatini and the Levi-Civita connection do not agree, the comparison of vectors in 
different points of the spacetime manifold will lead to different results, when performed 
with one connection or the other. In particular, one can think of configurations that would 
be static according to one connection, but not according to the other. A vector field 
$V^\mu(\lambda)$, representing some physical magnitude that evolves according to the equations 
of motion of the system with  initial conditions $V^\mu (\lambda_0)$, is said to be unchanged 
by the evolution of the system if its value $V^\mu (\lambda_1)$ for that magnitude
at a given time $\lambda_1$ is identical to the 
parallel transport of $V^\mu(\lambda_0)$ to $\lambda_1$. Now, let $V^\mu_{g}(\lambda)$ and  
$V^\mu_{\bGamma}(\lambda)$ be the results of parallel transport of $V^\mu(\lambda_0)$ according 
to the Levi-Civita and the Palatini connections respectively. As in general $V^\mu_{g}(\lambda)$ 
and $V^\mu_{\bGamma}(\lambda)$ will be different,  $V^\mu_{g}(\lambda_1) - V^\mu (\lambda_1)$ 
and $V^\mu_{\bGamma}(\lambda_1) - V^\mu (\lambda_1)$ will not be simultaneously zero.  The concept
of staticity is therefore as much related to the choice of connection, as it is to the dynamics 
of the system. So, in principle, one could think that the parallel transport should be 
observable.\footnote{\label{fWeyl}Weyl connections (introduced from different physical grounds, 
        see \cite{Enc}) also leaded to a parallel transport different to Levi-Civita's. 
        The detectability of this transport was essential in that theory, and its consequences 
        were criticised by Einstein. Notice, however, that Weyl tried to unify electromagnetism 
        and gravity, while Palatini connections emerge even when only the gravitational 
        interaction is taken into account. }
However, we believe that it is precisely the homothetic property of the Palatini connection
that turns the Palatini and the Levi-Civita parallel transports indistinguishable.  

As we have seen in Section \ref{properties}, the Palatini and the Levi-Civita 
transports are homothetic, such that in general $V^\mu_{g}(\lambda_1)$ and 
$V^\mu_{\bGamma}(\lambda_1)$ only differ by a (curve-dependent) overall factor, as shown in 
(\ref{homotheticfactor}). Configurations that are static according to one connection, would 
with respect to the other also appear static, upto a homothety. In other words, there are no
additional ``curvature effects'' associated to the Palatini transport, except for a change of
norm of the transported vector. Traditionally, the latter would be of course  interpreted as
non-staticity, but that is due to the fact that we are used to work with metric compatible
connections, where the invariance of the norm under parallel transport is guaranteed. The
real question is whether the norm of a parallel transported vector can be physically
detected in an experiment, or whether it is just a (useful) mathematical construction to
understand the theory. 

Traditionally, in General Relativity it assumed that one can define a unit measure in any
point, by defining it in one point and then transporting the measurement instrument
(say a rod), using the Levi-Civita connection. The rod is assumed to maintain its length,
as the different particles constituting the rod do not obey the geodesic deviation equation,
as they feel the electromagnetic or nuclear forces of the neighbouring particles, which,
except for the cases of extreme tidal forces, are much stronger than the curvature effects.
When on the other hand, the Palatini connection is used, it seems reasonable to argue that
the same physical arguments hold: the main forces acting on the individual particles of the
instruments are not the geometrical ones, but the ones created by neighbouring 
particles.\footnote{Note this important difference with Weyl's connections cited in 
             footnote \ref{fWeyl}.} 
Since the non-gravitational physics is unaffected as long as we are working with minimally 
coupled matter Lagrangian's, as we argued above, it seems reasonable to conclude that the 
homothetic character of the Palatini parallel transport, rather than an experimentally 
measurable property, becomes a mathematical issue to be taken into account when counting the 
descriptions of the same mensurable  system.

Finally, one could wonder whether the connection Palatini connection could give rise to 
observable quantum effects. A well-known example in quantum mechanics is the Aharonov-Bohm 
effect, where topologically distinct gauge fields $A_\mu$ give rise to physically different 
situations, even though they yield the same field strength tensor $F_\mn(A)$. However we 
believe that in our case 
there are no observable quantum effects associated to the vector field $\cA_\mu$: any field 
configuration for $\cA_\mu$ can be reabsorbed in a geodesic reparametrisation (\ref{G(lambda)}), 
independently of its field strengths $\cF_\mn(\cA)$ and independently of the topological class
the specific field configuration belongs to. In other words, in constrast to Maxwell theory, 
any choice of $\cF_\mn(\cA)$ is a gauge choice.

%%%%%%%%%%%%%%%%%%%%%%%%%%%%%%%%%%%%%%%%%%%%%%%%%%%%%%%%
\section{Interpretation and conclusions}
\label{conclusion}

The most general affine connection allowed by the Palatini formalism 
in the Einstein-Hilbert action (allowing also minimally coupled matter terms) is given by
the non-symmetric and non-metric compatible connection 
\be
\bGamma_\mn^\rho \ = \ \{_\mn^\rho\} \ + \ \cA_\mu \, \delta_\nu^\rho,
\ee 
with $\cA_\mu$ an arbitrary non-dynamical vector field. We have shown that the family of 
Palatini connections is furthermore unique in two ways, first of all because it is the only
connection (upto the exceptional case of footnote \ref{fn2})
that has the same pregeodesics 
as Levi-Civita and at the same time conserves the form of the Einstein equations and secondly
because it is the only connection that provides a parallel transport of vectors that is 
homothetic to the Levi-Civita transport along any curve. We have proven in the previous 
sections that this connection does not lead to physically observable effects at the level
of the Einstein equations or the trajectories of test particles and we have argued that
most likely neither it does when comparing the results of parallel transport of vectors.
As the Palatini connections are the unique ones that preserve this basic physics (Einstein 
equations and pregeodesics), the \textit{Palatini approach yields an exact variational 
characterization of such basic physics.}

So, if our interpretation is correct and
the Palatini connections indeed turn out to be unobservable in all physical situations, 
then this would hint to a kind of duality (beyond the gauge symmetry as stated in \cite{TW} or 
the invariance of the Einstein equation in \cite{Pons}) between spacetimes with different 
geometrical properties, as these would all display the same physics. 
In mathematical terms, this would mean that for every (pseudo-)Riemannian geometry that is a 
solution of the minimally coupled Einstein equations, there is a family of non-(pseudo-)Riemannian 
geometries that are mathematically distinct, but physically 
indistinguishable.

Probably the best way to see the geometrical origin of the Palatini connection is looking 
at the geodesic equation (\ref{LCgeodesic2}) and its functional (\ref{funcional}). When 
$\lambda$ is chosen to be an affine parametrisation (proper time, in physics language), 
then the geodesic equation acquires its standard form (\ref{LCgeodesic}). But when 
any other parametrisation is chosen, extra parametrisation-dependent terms appear in the 
equation for the pregeodesics. We have shown that these extra terms can be written as a 
scalar product $\dot x^\rho\cA_\rho$ between the velocity of the curve and some
specific vector field $\cA_\rho$,  independent of the curve,
which in turn can be combined with the Levi-Civita connection and be interpreted as a new, 
mathematically inequivalent connections $\bGamma_\mn^\rho$ .
It is therefore as if the Palatini formalism allows its users to freely choose the 
parametrisation of their geodesics, providing as solutions of the variational principle
those connections that under reparametrisation yield the standard Levi-Civita geodesics
with affine parametrisation (\ref{LCgeodesic}). However, notice that different non-symmetric
connections have the same geodesics (as the latter only depend on the symmetrised part of the
former) and then not all connections projectively 
related to Levi-Civita are allowed. Indeed, in order for the physics to be invariant, it is 
not enough that the new connection has the same pregeodesics, but also that the curvature 
tensors change in such a way that the Einstein equations are invariant. And as we have seen, 
the only connections that can do this, are precisely those selected by the Palatini 
formalism.\footnote{With the noteworthy exception of the connection 
           $\tilde \Gamma_\mn^\rho = \{_\mn^\rho \} + \cB_\nu \delta_\mu^\rho$ with 
           $\nabla_\mu \cB_\nu = \cB_\mu\cB_\nu$. We do not have a clear interpretation of 
           this specific case and leave the matter for futere investigation.}

Summing up the answer to our original problem is a bit subtler 
than expected: not only the Levi-Civita connection, but the entire family of Palatini 
connections are singled out by the variational principle and from a mathematical point of 
view, so there is no reason to assign a preferred status to Levi-Civita. However, \textit{since
  all Palatini connections lead to the same ``rough'' physics, the Levi-Civita connection
  has the virtue of being the simplest representative of a class of physically indistinguishable 
connections.}

We wish to emphasise that strictly speaking we can only make a hard statement about the
observability of the vector field $\cA_\mu$ in the realm of ``rough'' physics, not excluding
completely that the presence of $\cA_\mu$ might acquire a physical meaning in subtler
situations. However, if this were the case, we can not stop wondering why there is no
experimental evidence for the existence of this vector field in our universe. 
We leave these possible effects for future investigations.  

There are a number of ways the results of this letter can be extended. In the first place, 
it would be interesting to see whether the presence of $\cA_\mu$ could be detected in
more complicated situations, such as for example non-minimal couplings, higher curvature terms
or in a Jordan frame. Secondly, an obvious question is whether the Palatini connection 
as the most general solution to the variational principle is limited to the Einstein-Hilbert
actions, or whether it also appears in different theories. It is well known that the metric
and the Palatini formalisms are equivalent for Lovelock gravities, in the sense that 
the Levi-Civita connections appear as a solution to the Palatini equation for these theories.
However, %as far as we know, 
it is not clear whether it is a unique solution and, if not,
whether the Palatini connections appear also as an allowed solution by the variational principle 
(as far as we know, the only results in this direction appear in \cite{JS, Pons}).
Answering this question would also give hints on whether there are physically observable 
effects associated with the Palatini connections. Work on these topics by some of the authors
is in progress.

%%%%%%%%%%%%%%%%%%%%%%%%%%%%%%%%%%%%%%%%%%%%%%%%%%%%%%%%
%%%%%%%%%%%%%%%%%%%%%%%%%%%%%%%%%%%%%%%%%%%%%%%%%%%%%%%%
%%%%%%%%%%%%%%%%%%%%%%%%%%%%%%%%%%%%%%%%%%%%%%%%%%%%%%%%
%\newpage
\vspace{1cm}
\noindent
{\bf Acknowledgements}\\
We wish to thank Stanley Deser, Roberto Emparan, Pablo Galindo, Luis Garay, Roberto Giamb\'o, 
Friedrich W. Hehl, 
Olaf M\"uller, Josep Pons, Jos\'e M.M. Senovilla and Thomas Van Riet for useful discussions. 
The work of B.J. and J.A.O. was partially supported by the Junta de Andaluc\'{\i}a 
(FQM101) and the Universidad de Granada (PP2015-03). J.A.O. is also supported by a PhD contract 
of the Plan Propio de la Universidad 
de Granada. M.S. has been partially financed by the Spanish Ministry of Economy and 
Competitiveness and European Regional Development Fund (ERDF) through the project 
MTM2016-78807-C2-1-P.

%%%%%%%%%%%%%%%%%%%%%%%%%%%%%%%%%%%%%%%%%%%%%%%%%%%%%%%%%%
%\newpage

%%%%%%%%%%%%%%%%%%%%%%%%%%%%%%%%%%%%%%%%%%%%%%%%%%%%%%%
%\end{multicols}           %end multicolumns
%%%%%%%%%%%%%%%%%%%%%%%%%%%%%%%%%%%%%%%%%%%%%%%%%%%%%%%

\begin{thebibliography}{99}


\bibitem{We} S. Weinberg, {\it Gravitation and cosmology: principles and applications of the 
general theory of relativity}, John Wiley \& Sons Inc. N.Y. (1972). 



\bibitem{Palatini} A. Palatini, Rend. Circ. Mat. Palermo 43 (1919) 203. 

\bibitem{Hehl0} M. Blagojevi\'c and F.W. Hehl (Ed.), {\it Gauge theories of gravitation}, Imperial 
College Press, London (2013).

\bibitem{FFR} M. Ferraris, M. Francaviglia and C. Reina,
 General Relativity and Gravitation 14,  (1982) 243.
 
\bibitem{Ei} A. Einstein, Einheitliche Feldtheorie von Preuss.
Akad. Wiss. Berlin,  (1925) 414.

\bibitem{CMQ}  S. Cotsakis, J. Miritzis and L. Querella, J. Math. Phys. 40 (1999) 3063, 
                {\tt gr-qc/9712025}.

\bibitem{Querrella} L. Querella, {\it Variational Principles and Cosmological Models 
                in Higher-Order Gravity}, {\tt gr-qc/9902044}.

\bibitem{ABFO} G. Allemandi, A. Borowiec, M. Francaviglia and S. D. Odintsov, 
               Phys. Rev. D72 (2005) 063505, {\tt gr-qc/0504057}.

\bibitem{SL}  T.P. Sotiriou and S. Liberati, Annals Phys. 322 (2007) 935, {\tt gr-qc/0604006}.

\bibitem{TU}  V. Tapia and M. Ujevic, Class. Quant. Grav. 15 (1998) 3719, {\tt gr-qc/0605132}.  

\bibitem{LBM} B. Li, J.D. Barrow and D.F. Mota, Phys. Rev. D 76, 104047 (2007), 
                {\tt arXiv:0707.2664}.

\bibitem{IKPP} A. Iglesias, N. Kaloper, A. Padilla and M. Park, Phys. Rev. D76 (2007) 104001, 
                {\tt arXiv:0708.1163}.   

\bibitem{BD}  F. Bauer and D.A. Demir, Phys. Lett. B665: 222-226, 2008, {\tt arXiv:0803.2664}. 

\bibitem{CDV} S. Capozziello, F. Darabi and D. Vernieri, Mod. Phys. Lett. A26 (2011) 65-72, 
               {\tt  	arXiv:1006.0454 }.

\bibitem{Bauer}  F. Bauer, Class. Quant. Grav. 28 (2011) 225019, {\tt arXiv:1108.0875}.

\bibitem{CdL} S. Capozziello and M. De Laurentis, 
%{\it Extended Theories of Gravity}, 
Physics Reports 509 (2011) 167, 
               {\tt arXiv:1108.6266}.

\bibitem{Olmo} G. Olmo, {\it Introduction to Palatini theories of gravity and nonsingular 
                 cosmologies}, {\tt arXiv:1212.6393}.

\bibitem{ESJ} Q. Exirifard and M.M. Sheikh-Jabbari, Phys. Lett. B661: 158-161, 2008, 
              {\tt arXiv:0705.1879}.

\bibitem{BJB} 
    M. Borunda, B. Janssen and M. Bastero-Gil, JCAP 0811: 008, 2008,  {\tt arXiv:0804.4440};\\
    M. Bastero-Gil, M. Borunda and B. Janssen, {\it The Palatini formalism for higher-curvature 
            gravity theories}, in K.E. Kunze et al (ed.), {\it Physics and Mathematics of 
            Gravitation}, AIP Conference Proceedings 1122 (2009), 189-192, {\tt arXiv:0901.1590}.

\bibitem{DP} N. Dadhich and J.M. Pons, Phys. Lett. B 705 (2011) 139-142, {\tt arXiv:1012.1692}.


\bibitem{JS} B. Julia and S. Silva, Class. Quant. Grav. 15 (1998) 2173, {\tt gr-qc/9804029}.

\bibitem{Pons}  N.~Dadhich and J.M.~Pons,
 % ``On the equivalence of the Einstein-Hilbert and the 
%Einstein-Palatini formulations of general relativity for 
%an arbitrary connection,''
  Gen.\ Rel.\ Grav.\  {\bf 44}, (2012) 2337,
 % doi:10.1007/s10714-012-1393-9
  {\tt arXiv:1010.0869}.


\bibitem{TW} R.W. Tucker and C. Wang,  	Class. Quant. Grav. 12 (1995) 2587-2605, 
             {\tt gr-qc/9509011}.

\bibitem{Ortin} T. Ort\'{\i}n, {\it Gravity and Strings}, Cambridge University Press, 2004.

\bibitem{Deser} S. Deser, Class. Quant. Grav. 23 (2006) 5773, {\tt gr-qc/0606006}.
  
\bibitem{York} J.W. York, Phys. Rev. Lett. 28 (1972) 16 1082.

\bibitem{GH} G.W. Gibbons and S.W. Hawking, Phys. Rev. D. 15  (1977) 2752.

\bibitem{Hehl1} F.W. Hehl, P. von der Heyde,  G.D. Kerlick and J.M. Nester, 
                  Rev. Mod. Phys. 48 (1976) 393.


\bibitem{Spivak} M. Spivak, {\it A Comprehensive Introduction to Differential Geometry}, Vol II,
                 Ch. 6, Prop. 17,  Publish or Perish Inc, 1999.

\bibitem{SS} N.S. Swaminarayan and J.L. Sakko, J. Math. Phys. 24 (4) 1983.

\bibitem{Stephani} H. Stephani, D. Kramer, M. Maccallum, C. Hoensalaers and E. Herlt, 
                   {\it Exact solutions of Einstein's Field Equations},
                   Cambridge University Press, 2009.

\bibitem{Hall} G.S. Hall, {\it Symmetries and Curvature Structure in General Relativity},
                   World Scientific Lecture Notes in Physics: Volume 46, 2004.

\bibitem{Hehl2} F.W. Hehl, J.D. Mc Crea, E.W. Mielke and Y. Ne'eman,
  Phys. Rep. 258 (1995) 1.

\bibitem{BSS} O.F. Blanco, M. S\'anchez, J.M.M. Senovilla,  	
                    J. Eur. Math. Soc. 15 (2013) 595-634,
              {\tt  	arXiv:1101.5503}.

\bibitem{Enc} J.L. Bell and H. Kort\'e,  {\it Hermann Weyl}, Stanford Encyclopedia of 
                Philosophy (Summer 2015 Edition), Edward N. Zalta (ed.), \\
                {\tt http://plato.stanford.edu/archives/sum2015/entries/weyl/}

%\bibitem{St} W.O. Straub, {\it On the Failure of Weyl's 1918 %Theory},
%              {\tt http://vixra.org/abs/1401.0168}.


\end{thebibliography}
\end{document}